\documentclass[runningheads]{llncs}
\usepackage[T1]{fontenc}
\usepackage{graphicx}
\usepackage{makecell}
\usepackage{url}
\usepackage[percent]{overpic}
\usepackage{tikz}
\usepackage{tabularx}
\usepackage{booktabs}
\usepackage{todonotes}
\usepackage{listings}
\usepackage{subcaption}

\definecolor{codegray}{rgb}{0.95,0.95,0.95} % light gray background
\definecolor{codeblue}{rgb}{0,0,0.6}
\definecolor{codegreen}{rgb}{0,0.5,0}
\definecolor{codered}{rgb}{0.6,0,0}

\renewcommand{\figurename}{Figure}

\begin{document}
\title{Malleable Molecular Dynamics Simulations with GROMACS and DMR}
%
%\titlerunning{Abbreviated paper title}
% If the paper title is too long for the running head, you can set
% an abbreviated paper title here
%
\author{
  Petter Sandås\inst{1}\orcidID{0009-0009-2028-7719} \and
  Sergio Iserte\inst{1}\orcidID{0000-0003-3654-7924} \and
  Íñigo Aréjula-Aísa[1]\orcidID{0009-0006-7354-0915}\and
  Berk Hess\inst{2}\orcidID{0000-1111-2222-3333} \and
  Antonio J. Peña\orcidID{0000-0002-3575-4617}\inst{1}
}

\authorrunning{P. Sandås et al.}
% First names are abbreviated in the running head.
% If there are more than two authors, 'et al.' is used.
%
\institute{
  Barcelona Supercomputing Center (BSC)\\
  \email{\{petter.sandas, sergio.iserte, inigo.arejula, antonio.pena\}@bsc.es} \and
  KTH Royal Institute of Technology, Stockholm, Sweden\\
  \email{hess@kth.se}
}

\maketitle              % typeset the header of the contribution
%
%Up to 15 pages!

\begin{abstract}
Static resource allocations in high-performance computing (HPC) lead to inefficiencies for time-varying workloads, causing idle resources, queue delays, and higher node-hour costs.
The Dynamic Management of Resources (DMR) middleware enables MPI process malleability in Slurm via a simple API decoupled from scheduler internals.
In this work, we integrate DMR into the GROMACS molecular dynamics engine to obtain a malleable variant that can dynamically adapt its MPI process count by combining communication-efficiency-aware reconfiguration with GROMACS' native checkpoint/restart mechanism. 
We evaluate this design on the MareNostrum~5 supercomputer, comparing dynamic runs against static executions and quantifying reconfiguration overheads, time-to-solution, and node-hour savings for bursty GROMACS workloads.

\keywords{HPC \and Dynamic Resource Management \and MPI \and Malleability \and Molecular Dynamics}
\end{abstract}
\section{Introduction}
Modern high-performance computing (HPC) systems must cope with strongly time-varying application behavior, heterogeneous hardware, and highly contended shared infrastructures, yet most production workloads are still executed under static resource allocations decided at submission time. 
This rigidity often leads to idle resources or unnecessary queueing delays when job requirements change during execution, reducing overall system productivity and increasing node-hour consumption in large machines.

MPI process malleability has emerged as a promising mechanism to address these limitations by allowing running applications to expand or shrink their process sets at runtime in coordination with the resource manager~\cite{tarraf_malleability_2024}. 
The Dynamic Management of Resources (DMR) framework provides a middleware layer that decouples application-level malleability from scheduler internals, offering a small MPI-like API for initialization, readiness checking, and reconfiguration, and integrating with Slurm-based environments~\cite{iserte_resource_2025}.

GROMACS is a widely used molecular dynamics (MD) engine, heavily optimized for both CPUs and GPUs, and a key building block in biomolecular simulation workflows that are sensitive to both performance and turnaround time~\cite{VanDerSpoel2005}. 
Its traditional execution model assumes a fixed number of MPI ranks. 
However, large GROMACS campaigns typically consist of many independent simulations with heterogeneous and time-varying resource needs, making them a natural candidate for malleability to better exploit fluctuations in cluster load and improve node-hour efficiency in production settings. 

In this work, we integrate DMR into GROMACS to produce a malleable version that can dynamically adapt its process count at runtime. 
The integration is designed to be minimally invasive: DMR is initialized early in the GROMACS startup sequence, reconfiguration checks are inserted in the main molecular dynamics loop under a communication-efficiency strategy, and DMR is coordinated with the native checkpoint handler to trigger safe reconfigurations at neighbor-search steps. 

Specifically, this paper makes the following contributions:
\begin{itemize}
  \item We present a minimally invasive integration of the DMR framework into GROMACS, leveraging its existing checkpoint/restart facilities to enable MPI process malleability.
  \item We configure and evaluate a communication-efficiency-aware malleability policy for GROMACS on Marenostrum 5 supercomputer, quantifying reconfiguration overheads, time-to-solution, and node-hour consumption for bursty workloads compared to static executions.
\end{itemize}

The rest of the paper is structured as follows: Section~\ref{sec:background} provides background on GROMACS and DMR.  Section~\ref{sec:gromacs} describes the methodology for integrating DMR in GROMACS. Section~\ref{sec:evaluation} presents the evaluation results of dynamic executions compared to static ones.
Section~\ref{sec:related} reviews related work in malleability and HPC scientific applications.
Finally, Section~\ref{sec:conclusions} concludes the paper.

\section{Background}\label{sec:background}
The two main pieces of software, GROMACS and DMR, leveraged in this research are introduced in this section.

\subsection{GROMACS}

GROMACS (GROningen MAchine for Chemical Simulations) is a high-performance molecular dynamics (MD) package widely used for simulating biomolecular systems including proteins, lipids, and nucleic acids~\cite{VanDerSpoel2005}. GROMACS is highly relevant in the high-performance computing (HPC) community due to its exceptional performance optimization, excellent parallel scalability, and efficient utilization of modern heterogeneous computing architectures~\cite{Abraham2015}.

GROMACS is notable in HPC environments for several key reasons. The code is heavily optimized at the algorithmic and implementation levels, with carefully tuned neighbor searching algorithms and highly optimized inner loops that maximize floating-point throughput on modern CPUs. GROMACS exhibits excellent strong and weak scaling properties across thousands of CPU cores, making it ideal for large-scale cluster and supercomputer deployments~\cite{Pronk2013}.

The package provides full support for GPU acceleration through CUDA and OpenCL. This heterogeneous computing capability is critical for maximizing throughput on modern HPC systems equipped with GPU accelerators. GROMACS employs efficient domain decomposition parallelization, distributing spatial domains across MPI ranks with optimized load balancing to minimize communication overhead~\cite{Hess2008}.

These characteristics make GROMACS one of the most widely deployed molecular dynamics packages on top-tier supercomputers worldwide, routinely appearing in HPC benchmarking studies and demonstrating sustained performance on leading embarrassingly parallel workloads.

\subsection{DMR}

The Dynamic Management of Resources (DMR, pronounced ``dimmer'') framework is a middleware layer designed to enable process malleability and dynamic resource management for MPI applications in HPC environments~\cite{iserte_agut_high-throughput_2018}. It serves as an orchestration bridge between the scientific application, the MPI runtime (i.e., Open MPI/PRRTE~\cite{gabriel_2004}), performance monitoring tools (i.e., TALP~\cite{lopez_talp_2021}), and the underlying RMS (i.e., Slurm \cite{yoo_slurm_2003}). The primary objective of DMR is to allow running jobs to dynamically acquire, release, or reconfigure compute resources at runtime, enabling applications to expand or shrink their resource footprint transparently to improve system utilization and responsiveness.

Internally, DMR follows a state-driven architecture, which is exposed to developers through a reduced set of API routines~\cite{iserte_dmrlib_2020}:
\begin{itemize}
    \item \texttt{dmr\_init}: Initializes the dynamic environment, configures communication with the RMS, and prepares internal data structures.
    \item \texttt{dmr\_check}: Invoked at synchronization points to indicate readiness for resizing; it evaluates policies and system feedback to decide whether a reconfiguration is necessary.
    \item \texttt{dmr\_reconfigure}: Executes the selected reconfiguration and coordinates with the MPI runtime to spawn or terminate processes as required.
    \item \texttt{dmr\_finalize}: Handles the final cleanup of the management layer and releases resources.
\end{itemize}

A significant advancement in the framework is the introduction of a non-invasive methodology specifically tailored for production HPC clusters. While originally, it was often relied on customized or "invasive" scheduler modifications ---such as Slurm4DMR--- in controlled environments, with the new \textit{DMR@Jobs} feature, malleable applications are able to directly interact with standard, unmodified RMS deployments through user-level API calls. 
This approach orchestrates MPI processes across multiple independent jobs to achieve expansion and shrinkage without requiring administrative privileges. Furthermore, DMR provides a unified programming model that supports both \textbf{in-memory data redistribution} via MPI and \textbf{checkpoint-restart (C/R)} mechanisms, allowing it to adapt to diverse scientific codes and existing application-level checkpointing logic.

The framework supports sophisticated reconfiguration strategies, such as the \texttt{CE\_POLICY}, which utilizes performance metrics to adjust allocations based on a target communication efficiency. By allowing applications to continue executing while resource requests are pending, the framework effectively hides resource acquisition latency in heavily contended production systems. This makes DMR a robust solution for bridging the gap between experimental malleability prototypes and operational supercomputing environments~\cite{iserte_mpi_2025}.

\section{GROMACS Malleable}\label{sec:gromacs}
This section describes how GROMACS has been adapted to support malleability through integration with the DMR framework. The modifications were designed to be minimally invasive, leveraging GROMACS' existing checkpointing mechanism to enable dynamic resource management without requiring extensive changes to the core simulation logic.

GROMACS follows a traditional time step-based approach with MPI parallelization, where the simulation domain is decomposed spatially and distributed across MPI ranks. This design allows for efficient scaling but does not natively support dynamic changes in the number of processes during execution. To enable malleability, we have integrated DMR into key points of the GROMACS codebase to allow for runtime reconfiguration while ensuring that the simulation state remains consistent and recoverable.

Data redistribution is not explicitly implemented in this integration, as the checkpoint-restart mechanism of GROMACS is leveraged to handle state recovery and continuation across reconfigurations. This approach simplifies the implementation while still providing the necessary functionality for dynamic resource management.

\subsection{Initialization}

The DMR initialization routine is invoked early in the GROMACS execution flow to configure the necessary environment for dynamic resource management.

\begin{figure}
\begin{lstlisting}[language=C, caption=Initialization of DMR within GROMACS., label=lst:dmr-init, captionpos=b]
void init(int* argc, char*** argv) {
    ...
    DMR_AUTO(dmr_init(*dmr_argc, *dmr_argv), (void)NULL, (void)NULL, (void)NULL);
}
\end{lstlisting}
\vspace{-0.8cm}
\end{figure}

Listing~\ref{lst:dmr-init} illustrates the invocation of \texttt{dmr\_init} within the GROMACS initialization sequence. The arguments passed to \texttt{dmr\_init} are derived from the command-line arguments provided to GROMACS, allowing DMR to inspect and adjust them as needed to ensure compatibility with the malleable execution model.

The data redistribution arguments are intentionally left unspecified (i.e., set to \texttt{NULL}). This design choice reflects the fact that the proposed technique leverages GROMACS' native checkpointing mechanism for state recovery and continuation. As a result, no explicit data redistribution routines are required within DMR: the checkpoint-based restart mechanism, as handled internally by GROMACS, is sufficient to ensure consistency across reconfigurations.

With more details, during the initialization phase, DMR inspects the runtime arguments passed to GROMACS to ensure that they are compatible with the malleable execution model. If certain arguments are missing, such as \texttt{-cpi <filename>} for checkpointing or \texttt{-append} for output management, DMR programmatically inserts them to enforce the necessary conditions for dynamic reconfiguration. This ensures that GROMACS can produce consistent checkpoints and manage output files appropriately across reconfigurations.

\subsection{Reconfiguration}

The synchronization point for reconfiguration is integrated into the main simulation loop of GROMACS, as shown in Listing~\ref{lst:dmr-reconf}. This loop iteratively advances the simulation through time steps, and it provides a natural point at which DMR can evaluate whether a reconfiguration should occur based on the current state of the application and the availability of resources.

\begin{figure}
\begin{lstlisting}[language=C, caption=Reconfiguration logic within GROMACS., label=lst:dmr-reconf, captionpos=b]
void gmx::LegacySimulator::do_md()
{
    ...
    /* Loop over MD steps */
    ...
    if(dmr_check(CE_POLICY))
        dmr_ready_reconfig = true;
    ...
    if(dmr_ready_reconfig) {
        dmr_reconfigure();
    ...
}
\end{lstlisting}
\vspace{-0.8cm}
\end{figure}

The integration logic relies on a boolean state machine to detect whether DMR has determined that a reconfiguration is necessary, based on the active policy and resource availability. 
In the positive case (line~10), a flag is set to indicate that the application will be reconfigured in line~14. 
This preparation involves coordinating with GROMACS' checkpointing mechanism to ensure that a consistent checkpoint is produced before invoking the reconfiguration. 

Following this, \texttt{dmr\_reconfigure} is invoked, and GROMACS is allowed to terminate through its normal shutdown pathway. Process reallocation and restart are managed invisibly by DMR.

In this paper, the \texttt{CE\_POLICY} is used as the policy for triggering reconfiguration. This policy evaluates the communication efficiency of the application and determines whether a reconfiguration is warranted to optimize the resource utilization by approching a target efficiency~\cite{iserte_parallel_2024}.

Notably, the \texttt{DMR\_AUTO} macro is not used in this context, since its default behavior would early-terminate the execution before the opportunity to produce a consistent checkpoint because of GROMACS structure. Instead, the logic is implemented manually to allow for a safe checkpointing point to be reached before invoking the reconfiguration.

\subsection{Checkpoint Coordination}

For the integration of DMR to function correctly, it is crucial that GROMACS produces checkpoints at appropriate times during the simulation, particularly when a reconfiguration is triggered. This ensures that the application state can be reliably saved and restored across changes in the resource allocation.

We have introduced in \texttt{src/gromacs/mdlib/checkpointhandler.cpp} additional modifications within the function \texttt{decideIfCheckpointingThisStepImpl}, which governs when checkpointing should occur.

In a standard GROMACS execution, it interprets external signals (e.g., \texttt{SIGTERM}) to trigger checkpointing. Upon receiving such a signal, GROMACS typically defers checkpoint creation until a neighbor-search step is reached, ensuring that the simulation state is left in a consistent and well-defined configuration.

To integrate DMR into this mechanism, a call to \texttt{dmr\_get\_last\_action} is introduced. If the returned value indicates that the most recent DMR action is a reconfiguration, this condition is treated equivalently to an external checkpointing signal. Consequently, checkpointing is deferred in the same manner, i.e., until the next neighbor-search step is reached.

\section{Performance Evaluation}\label{sec:evaluation}

In our experiments, we demonstrate how in-production execution can save a non-negligible amount of compute node-hours compared to executions in exclusive reservations of resources.

\subsection{Setup Configuration}

\subsubsection{Infrastructure}
All the experiments are executed on MareNostrum 5 (MN5), a pre-exascale system integrated into the EuroHPC-JU infrastructure\footnote{\url{https://bsc.es/marenostrum/marenostrum-5}}. 
Particularly, jobs are submitted to its general-purpose partition (GPP)\footnote{\url{https://www.bsc.es/supportkc/docs/MareNostrum5/overview/#marenostrum-5-gpp-general-purpose-partition}}, which comprises 6,192 Intel Sapphire Rapids-based nodes, each equipped with two Intel Xeon Platinum 8480+ processors (56 cores at 2~GHz, 112 cores total) and 256~GB of DDR5 memory, interconnected by a 100~Gbit/s ConnectX-7 NDR200 InfiniBand network and managed by Slurm~24.11.6.

Open MPI~5.1.0a1, PRRTE~5.0.0a1, OpenPMIX~7.0.0a1, DLB~3.5.0, and DMR~2.0.1, conform the software stack used for the experiments.

Our experiments are performed in a controlled environment with Slurm4DMR, which is a customized version of Slurm that includes the necessary modifications to support DMR's malleability features. This environment is copnfigured on a dedicated set of nodes within MN5, ensuring that the malleable jobs can be executed without interference from other users or workloads.

The experiments are scoped within 32 nodes for up to 2 hours, which fits in the QoS of the debug partition in MN5. Notice that one of the alllocated nodes is reserved to serve as the Slurm controller node, while the remaining 31 nodes act as actual compute nodes.

\subsubsection{Application}
The malleable version of GROMACS designed in this paper is based on its 2025.4 version and evaluated with the mdp input \textit{stmv}~\cite{szilard_pall_2020_3893789}, the Satellite Tobacco Mosaic Virus solvated in TIP3P using cubic box and the CHARMM27 force field. It is configured with $1,066,628$ atoms, $2$ fs time-step, $1.2$ nm cut-offs, h-bond constraints, $0.15$ nm PME grid spacing, NVT ensemble, and limited to $5,000$ time steps.
The application is instantiated with one MPI rank per node and 112 OpenMP threads per rank.

\subsubsection{Malleability}
Malleability uses \texttt{CE\_POLICY}, which monitors communication efficiency (CE) ---the fraction of execution time spent in computation versus communication--- via the TALP monitor. Ranks adjust linearly with deviation from a 95\% target CE; high CE (>95\%) triggers expansion for faster execution, low CE (<95\%) triggers shrinkage to cut contention.

Reconfiguration decisions are made every 500 timesteps; DMR implements a feature known as the inhibitor, which enforces a configurable timestep-based delay among reconfigurations.

Malleable jobs are submitted with a node range from 1 to 12 (\texttt{-N1-12}), corresponding in turn to the lower and upper bounds of the reconfigurations, which means that the job can shrink down to 1 rank or expand up to 12 ranks.

\subsubsection{Workload}
The malleable GROMACS application is instantiated through Slurm jobs submitted to request computational resources.
Each experiment generates and submits a workload of 10 such jobs.

As common in MPI malleability evaluations to maximize system stress~\cite{iserte_dmrlib_2020}, jobs have virtually zero inter-arrival time, yielding a burst submission of all 10 jobs.

\iffalse
\subsubsection{Environments}
We evaluate the malleable version of GROMACS in two different environments: 1) a \textit{controlled} environment with Slurm4DMR and 2) a \textit{production} environment with DMR@Jobs.

Both environments are deployed on MN5, and they are configured with the same malleability settings and workload characteristics. The key difference between them is the mechanism through which malleability is enabled: Slurm4DMR in the controlled environment requires a malleability-enabled Slurm version running on a set of nodes that are exclusively allocated for the experiments; while DMR@Jobs in the production environment operates with the standard MN5 Slurm deployment, allowing the malleable jobs to be scheduled alongside other users and workloads in the shared cluster.

The experiments are scoped within 32 nodes for up to 2 hours, which fits in the QoS of the debug partition in MN5. Notice that a node from the allocation acts as the controller, while the remaining 31 nodes act as actual compute nodes.
\fi

\subsection{Results}
In this section, we present the results of the performance evaluation of the malleable version of GROMACS within different workloads.

\subsubsection{Baseline}
In order to establish a baseline for comparison, we execute the same workload of ten GROMACS jobs in a static configuration, where each job is allocated a fixed number of nodes (either 2 or 12) for the entire duration of its execution. This allows us to quantify the benefits of malleability by comparing the behavior of the malleable executions against these static baselines.

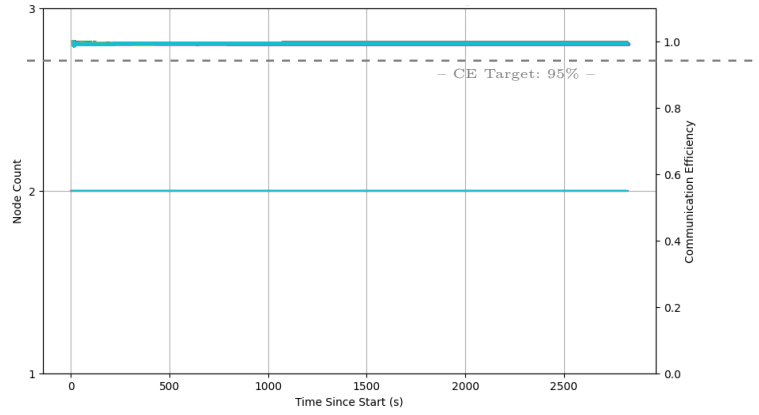
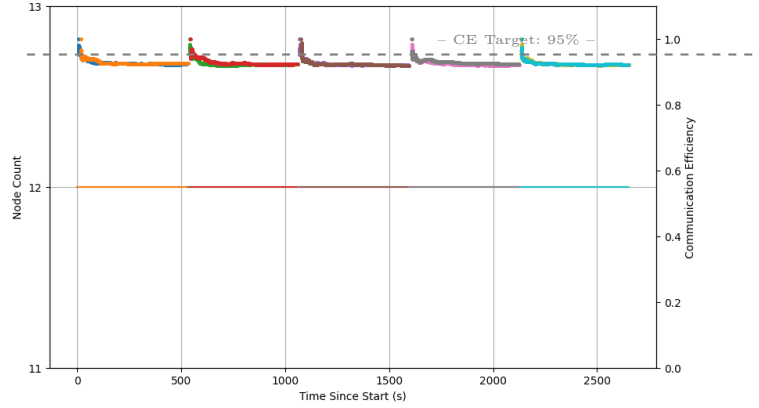
\begin{figure}
    \centering
    % Subfigure (a)
    \begin{subfigure}[b]{\linewidth}
        \centering
        \begin{overpic}[clip,width=0.75\linewidth,trim={0.1cm 11.25cm 0.1cm 0.71cm}]{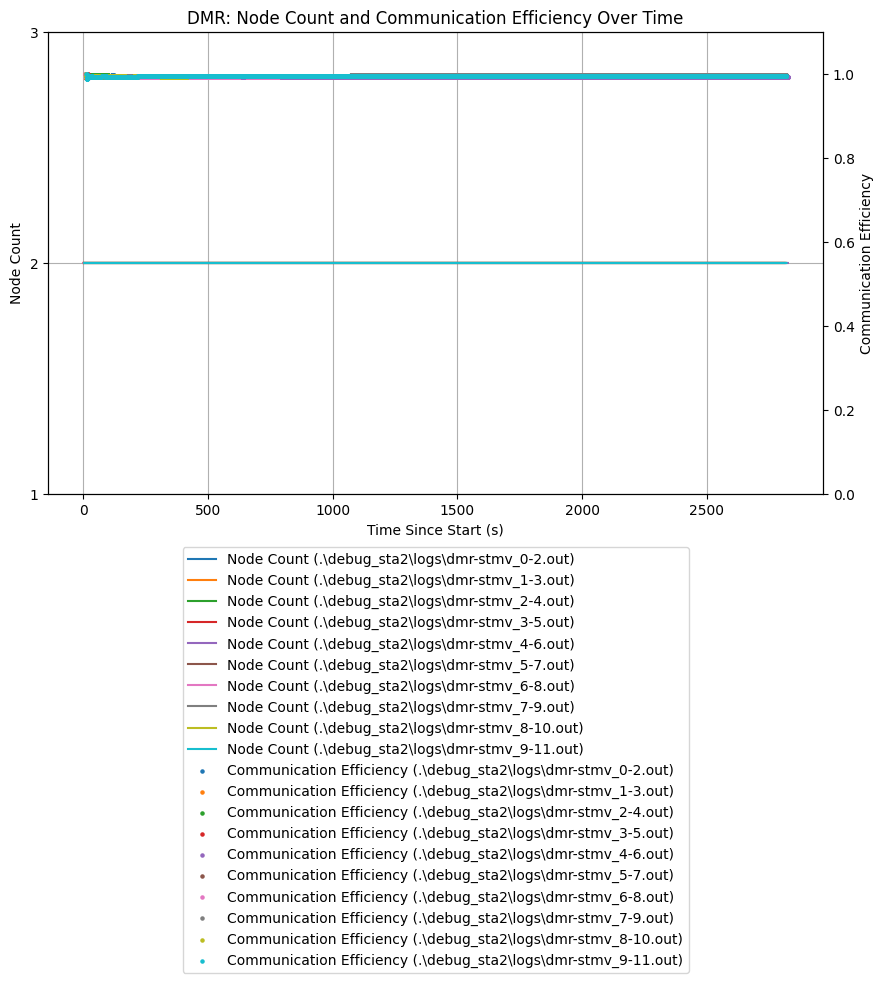}
            \put(2.5,50.375){\begin{tikzpicture}
                \draw[gray, dashed, thick] (0,0) -- (0.8\linewidth,0);
            \end{tikzpicture}}
            \put(62,48){\color{gray}\tiny -- CE Target: 95\% --}
        \end{overpic}
        \caption{Execution with 2 nodes per job.}
        \label{fig:sta2}
    \end{subfigure}
    \hfill
    \\
    \vspace{0.5cm}
    % Subfigure (b)
    \begin{subfigure}[b]{\linewidth}
        \centering
        \begin{overpic}[clip,width=0.75\linewidth,trim={0.1cm 11.25cm 0.1cm 0.71cm}]{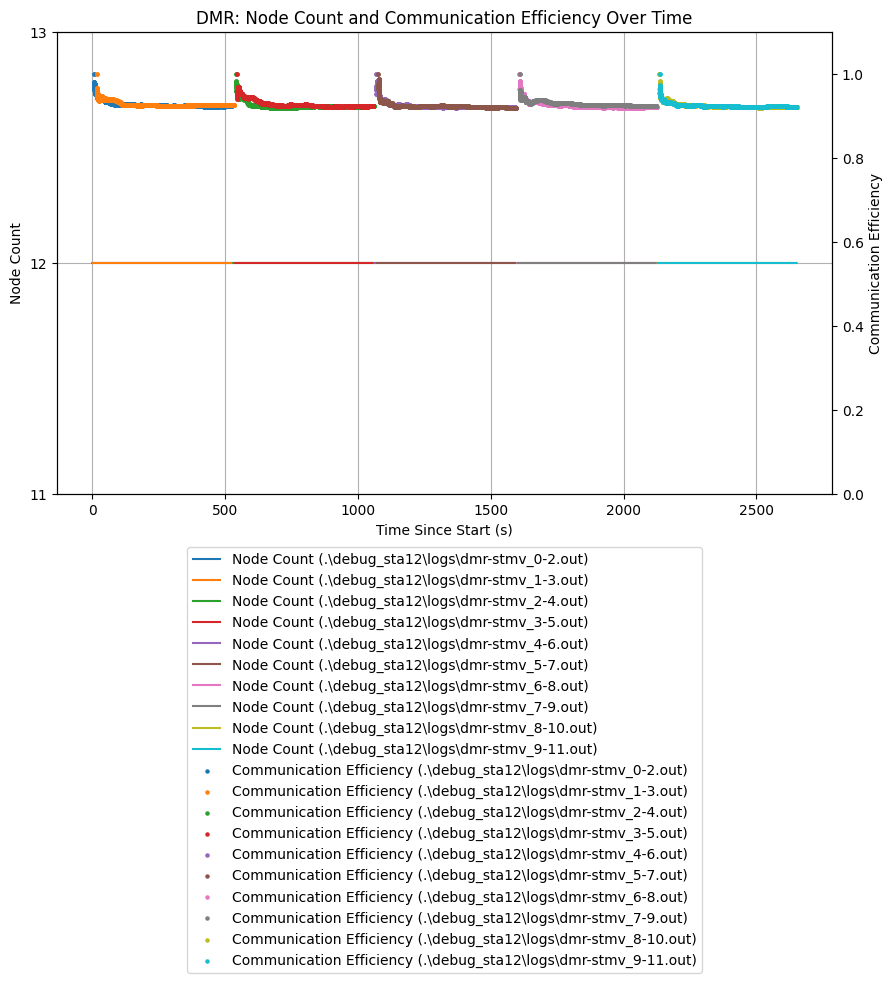}
            \put(2.5,50.375){\begin{tikzpicture}
                \draw[gray, dashed, thick] (0,0) -- (0.8\linewidth,0);
            \end{tikzpicture}}
            \put(62,52){\color{gray}\tiny -- CE Target: 95\% --}
        \end{overpic}
        \caption{Execution with 12 nodes per job.}
        \label{fig:sta12}
    \end{subfigure}
    % Overall caption
    \caption{Execution of the workload in the static configuration.}
    \label{fig:static}
\end{figure}

Figure~\ref{fig:static} shows the execution of the workload in the static configurations with 2 and 12 nodes, respectively. 
The charts illustrate the allocated nodes (left y-axis) over the execution time (x-axis) for each job, where each color represents one of the ten jobs.
Furthermore, the y-axis on the right side shows the CE of each job (in the same color scheme as the allocated nodes) over time.

The 2-node static configuration (Figure~\ref{fig:sta2}) allocates two nodes to each job for the entirety of its execution, maintaining stable CE levels close to the maximum efficiency. All jobs are submitted at the same time and execute in parallel; therefore, the total runtime approximately corresponds to the individual execution time of a single job. In total, 20 nodes are allocated, yielding a makespan of $2,825$ seconds.
In contrast, the 12-node static configuration (Figure~\ref{fig:sta12}) shows that all jobs are allocated 12 nodes throughout their execution, exhibiting greater fluctuations in CE levels. The increased availability of resources leads to a shorter individual execution time for each job, but the CE levels are generally lower than in the 2-node configuration due to the increased communication overhead among processes. All the jobs are submitted simultaneously, but only two can run simultaneously ($12+12$) due to the node partition limit of 31.
In this case, 24 nodes remain allocated during the makespan of $2,652$ seconds.

\subsubsection{Dynamic Resource Management}
\begin{figure}
    \centering
    \begin{overpic}[clip,width=0.999\linewidth,trim={0.1cm 11.25cm 0.1cm 0.75cm}]{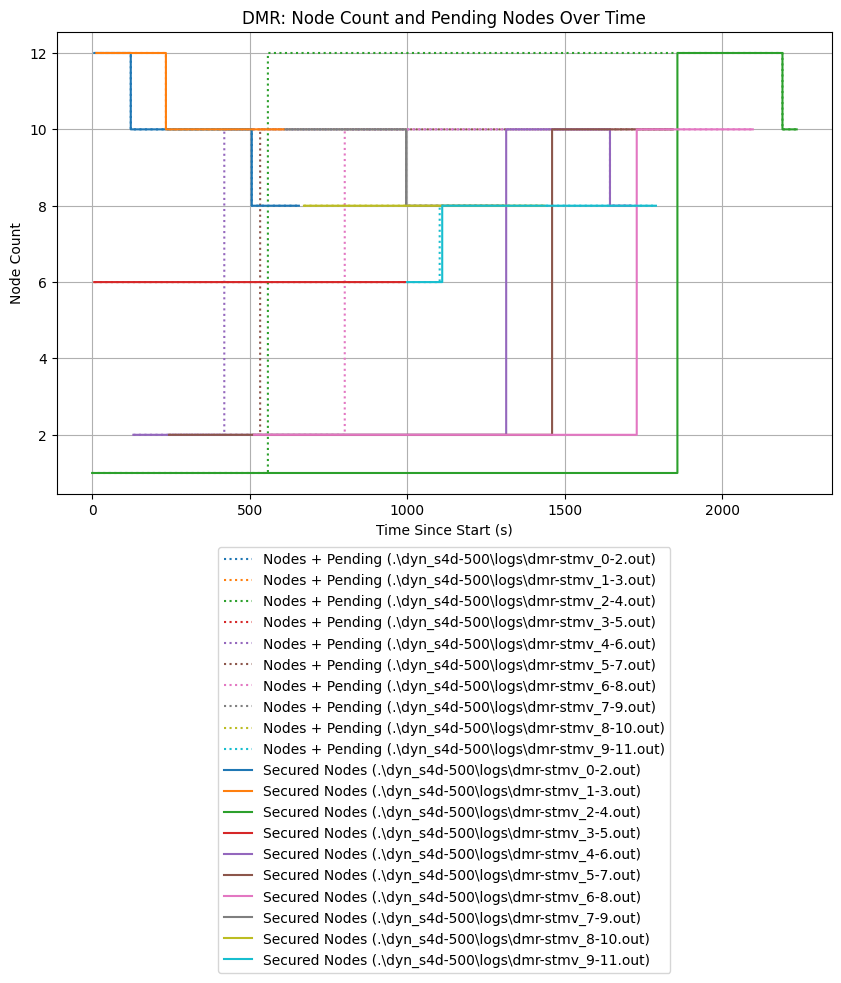}
        %put(x from left, y from bottom){text}
        \put(8,55){1st}
        \put(20,55){2nd}
        \put(8,31.5){3rd}
        \put(8,8.75){4th}
        \put(12,13){5th}
        \put(18,13){6th}
        \put(28,13){7th}
        \put(32,50){8th}
        \put(35,41){9th}
        \put(45,31.5){10th}
    \end{overpic}
    \caption{Execution of the workload with malleable jobs, showing the allocated nodes. The dotted lines indicate the scheduled expansions that are pending until resources are available and its next synchronization point is reached.}
    \label{fig:dyn-s4d-pend}
      \vspace{-0.4cm}
\end{figure}
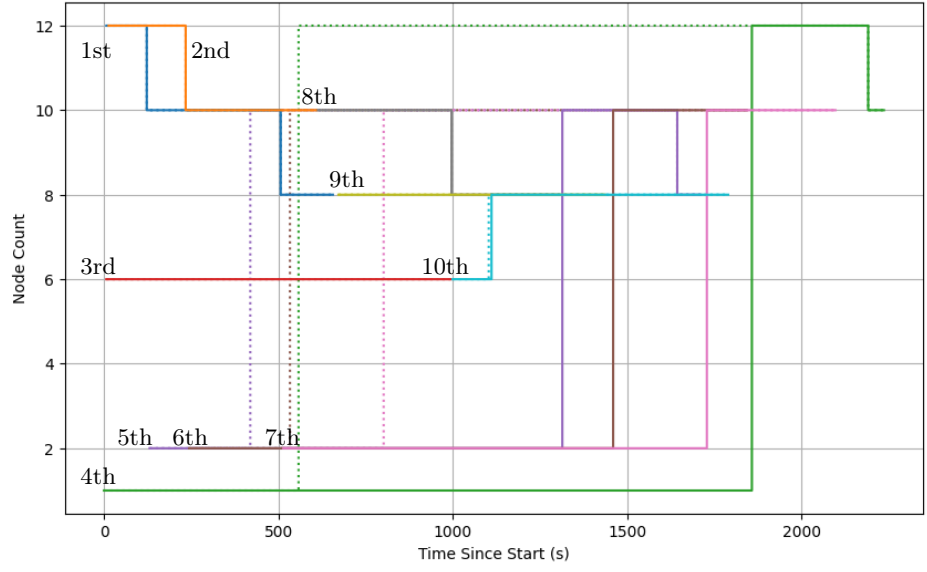

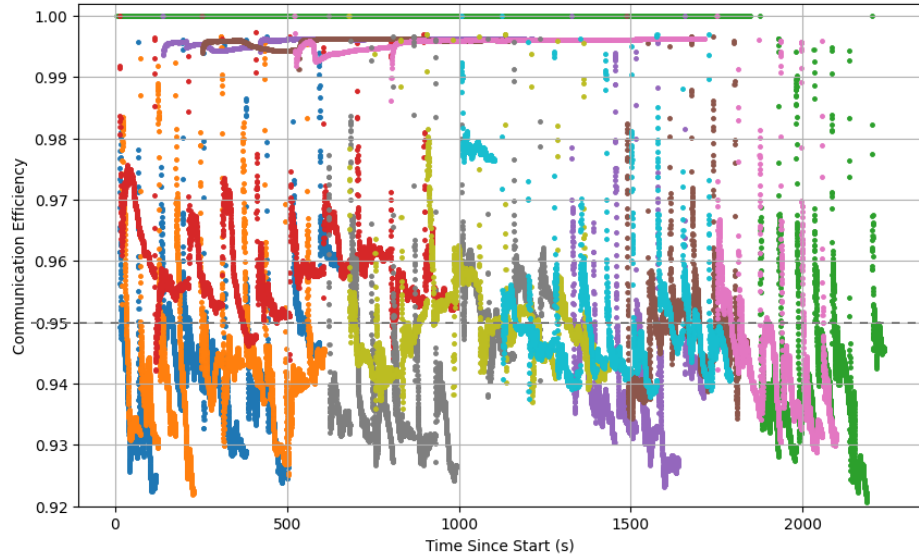
\begin{figure}
    \centering
    \begin{overpic}[clip,width=0.999\linewidth,trim={0.1cm 5.75cm 0.1cm 0.71cm}]{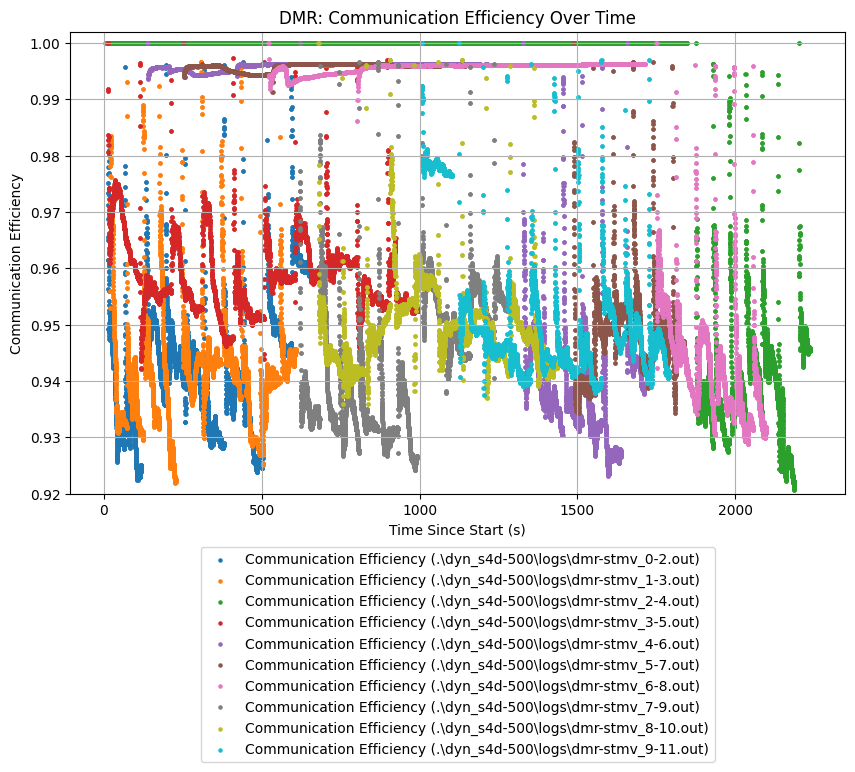}
        \put(2.5,25.9){\begin{tikzpicture}
            \draw[gray, dashed, thick] (0,0) -- (\linewidth,0);
        \end{tikzpicture}}
        %\put(62,52){\color{gray}\tiny -- CE Target: 95\% --}
    \end{overpic}

    \caption{Communication efficiency of the malleable jobs over time.}
    \label{fig:dyn-s4d-ce}
\end{figure}

\begin{figure}
    \centering
    \includegraphics[clip,width=0.999\linewidth,trim={0.1cm 0.25cm 0.1cm 0.71cm}]{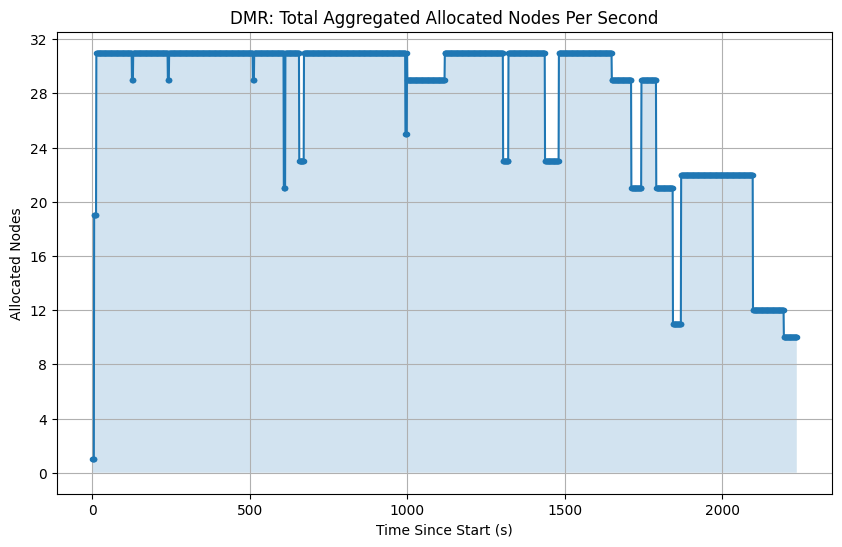}
    \caption{Overall allocated nodes over time for the malleable workload execution.}
    \label{fig:dyn-s4d-alloc}
\end{figure}

When resubmitting the same workload with malleable jobs, we observe a different behavior in terms of resource allocation and CE levels, as shown in figures~\ref{fig:dyn-s4d-pend},~\ref{fig:dyn-s4d-ce}, and~\ref{fig:dyn-s4d-alloc}.

Particularly, as shown in Figure~\ref{fig:dyn-s4d-pend}, the jobs are not all running simultaneously from the beginning of the execution; they are scheduled and started at different times as resources become available. This leads to a more staggered execution pattern, where some jobs may start later than others. However, the jobs can also finish earlier due to the dynamic allocation of resources.
Specifically, the first four jobs are started with 12, 12, 6, and 1 nodes, respectively, allocating the 31 available nodes.
As the execution progresses, the 12-node jobs are shrunk to 10, according to the DMR policy, which detects that their CE is lower than the target threshold in Figure~\ref{fig:dyn-s4d-ce}.
With the first nodes released, the fifth job is initiated with 2 nodes, and likewise the sixth job is assigned 2 nodes shortly after.
The dotted lines in the CE chart (Figure~\ref{fig:dyn-s4d-pend}) indicate the time in which a job is scheduled to be expanded, but in which the expansion cannot occur until both \textbf{resources are available and the next synchronization point is reached}.
For example, the fifth job is scheduled to be expanded at around 450 seconds, but it cannot be expanded until around $1,300$ seconds have elapsed.

Figure~\ref{fig:dyn-s4d-ce} plots CE over time for malleable jobs.
Jobs 3--5 start with higher, more stable CE values due to fewer initial nodes.
Higher node counts yield lower, more variable CE from greater inter-process communication overhead.
As in the 12-node static case, CE spikes at the start (or after a reconfiguration), then it stabilizes as averages accumulate.
Reconfigurations are configured asynchronously, so application execution continues while new resources are awaited, resulting in only amortized overhead.
Once resources are granted, reconfiguration awaits the next synchronization point.

Table~\ref{tab:overhead} presents the reconfiguration times.
Expansions are slower than shrinks due to these asynchronous waits, an effect pronounced at low node counts where slower iterations delay the \texttt{dmr\_check} call~\cite{doi:10.1177/10943420231176527}.

Figure~\ref{fig:dyn-s4d-alloc} shows total allocated nodes over time ---the sum across all jobs at each timestep.
At workload start, all 31 compute nodes remain allocated.
As jobs complete and queue length drops, resources are released.
However, availability does not guarantee expansion: if CE already exceeds the target threshold, new resources are not requested.

Table~\ref{tab:workload-stats} compares the three workloads by total execution time (\textit{makespan}, column 2, in seconds) and computational cost.
The latter is computed via trapezoidal integration of the allocated-node profile  over time (figures~\ref{fig:sta2},~\ref{fig:sta12}, and~\ref{fig:dyn-s4d-alloc}), reported in node-hours (\textit{net cost}, column 3).

The dynamic workload finishes fastest, since expansions accelerate jobs ---even with $184.37$\,s total overhead from 11 reconfigurations ($8.24\%$ of makespan).
Conversely, the 2-node run is the slowest but cheapest, thanks to higher efficiency from reduced inter-node communication.

\begin{table}
    \vspace{-0.4cm}
\centering
\setlength{\tabcolsep}{22pt}
\caption{Aggregated reconfiguration times.}
\label{tab:overhead}
    \begin{tabularx}{\linewidth}{r c l}
        \toprule
        \textbf{Category} & \textbf{Count} & \textbf{Time} \\
        \midrule
            All     & 11 & $16.76 \pm 10.54$ s \;\; [$7.83$ -- $42.44$] s \\
            Expands & 5  & $25.55 \pm 9.99$ s \;\; [$15.40$ -- $42.44$] s \\
            Shrinks & 6  & $9.43 \pm 1.63$ s \;\; [$7.83$ -- $12.34$] s \\
        \bottomrule
    \end{tabularx}
        \vspace{-0.4cm}
\end{table}

\begin{table}
        \vspace{-0.4cm}
  \centering
  \setlength{\tabcolsep}{10pt}
  \caption{Aggregated statistics per workload.}
  \label{tab:workload-stats}
  \begin{tabularx}{\linewidth}{Xccc}
    \toprule
    & \textbf{Makespan} & \textbf{Net cost} & \textbf{Total cost} \\
    \midrule
    Static 2-node jobs & $2,825$ s & $15.63$ n-h& $25.11$ n-h\\
    Static 12-node jobs & $2,652$ s & $17.53$ n-h& $23.57$ n-h\\
    Dynamic [1--12]-node jobs  & $2,236$  s & $17.15$ n-h & $19.87$ n-h\\
    \bottomrule
  \end{tabularx}
      \vspace{-0.4cm}
\end{table}

Nevertheless, workloads run in \textit{workflow mode} with pre-allocated resources: 32 nodes total, one dedicated to the Slurm controller.
The per-experiment total cost appears in column 4.
With 31 compute nodes, the dynamic case adapts jobs to availability (Figure~\ref{fig:dyn-s4d-alloc}), guided by the CE policy.
Static cases cannot automatically be adapted to the available resources.
For instance, in the 2-node case, the ten jobs are running concurrently, using 20 nodes out of 31 (see Figure~\ref{fig:sta2}), while in the 12-node case, six groups of two concurrent jobs run sequentially allocating 24 nodes (see Figure~\ref{fig:sta12}).

\section{Related Work}\label{sec:related}
Several research efforts have focused on developing malleable versions of full scientific applications to evaluate dynamic resizing frameworks. Early evaluations often utilized the NAS Parallel Benchmarks (NPB), including kernels such as Integer Sort (IS), Conjugate Gradient (CG), Fourier Transform (FT), and the pseudo-application LU, particularly to test the ReSHAPE framework~\cite{sudarsan-nas}. 

The CG method and Jacobi solver have also been implemented on top of Flex-MPI, along with Epigraph, which represents a more complex high-performance computing (HPC) application~\cite{flexmpi}. In the domain of molecular dynamics, the mini-application LeanMD was adapted using the Charm++ runtime to implement simplified force calculations~\cite{leanmd}. Furthermore, LAMMPS, a highly complex classical molecular dynamics code, was developed into a malleable version by leveraging its checkpoint/restart (C/R) capabilities under the direction of ReSHAPE~\cite{sudarsan-lammps}.

A malleable version of MPDATA, a fundamental component of the EULAG multiscale fluid solver, responsible for computing the advection of a non-diffusive quantity in a flow field by iterative time-stepping, has been implemented with DMR API with~\cite{iserte_study_2020} and without~\cite{iserte_resource_2025} GPU acceleration.

Unlike many previous efforts that focused on regular iterative patterns, HPG-aligner, a bioinformatics tool for RNA sequence mapping, represents a non-iterative producer-consumer application with irregular communication patterns and complex data structures implemented with DMR API~\cite{iserte_dynamic_2018}.

Alya is a high performance computational mechanics code to solve coupled engineering problems which also adopted C/R-based malleability and communication efficiency-based reconfigurations with DMR~\cite{iserte_malleable_2025}.

To the best of our knowledge, this is the first work to develop and evaluate a malleable version of GROMACS, demonstrating its viability and benefits for dynamic resource management in HPC environments.

\section{Conclusion}\label{sec:conclusions}
This work has presented a minimally invasive integration of the Dynamic Management of Resources (DMR) framework into GROMACS, enabling MPI process malleability by reusing the existing checkpoint/restart mechanism and a small set of well-placed integration points in the initialization, main loop, and checkpoint handler. 

Our evaluation has compared a malleable GROMACS configuration against static workloads. 
The dynamic run achieves the shortest makespan because expansions leverage available resources and accelerate individual jobs, even though reconfigurations introduce an aggregate overhead of only a few percent of the total execution time. 
These results confirm that communication-efficiency-guided malleability can successfully trade off time-to-solution and resource usage scenarios, and that DMR-enabled GROMACS can exploit resources more effectively than static counterparts.

Beyond current x86-based clusters, the European PILOT (EUPilot) project aims to demonstrate a pre-exascale European accelerator platform based on long-vector RISC-V technology, co-designing hardware, system software, and flagship scientific applications~\cite{eupilot2021eu}. 
Within this context, a RISC-V vector version of GROMACS is being developed as part of the EUPilot application portfolio, and is expected to become dynamically malleable through DMR, enabling the same communication-efficiency-driven reconfiguration strategies on future European RISC-V accelerators as those explored here on MN5. 
Our work thus provides an essential step towards a unified malleability story for GROMACS across current x86 supercomputers and forthcoming RISC-V-based demonstrators. 

\section*{Acknowledgments}
This project is co-funded by the project The European PILOT, which has received funding from the European High-Performance Computing Joint Undertaking (JU) under grant agreements No. 101034126 and No. PCI2021-122090-2A under the MCIN/AEI and the EU NextGenerationEU/PRTR.

BSC researchers are supported by the Ministerio para la Transformación Digital y de la función pública, within the framework of the Plan de Recuperación Transformación y Resiliencia, and by the European Union – NextGenerationEU. 

Antonio J. Peña was partially supported by the Ramón y Cajal fellowship RYC2020-030054-I funded by MCIN\slash AEI\slash 10.13039\slash 501100011033 and by ``ESF Investing in your future''.

The views and opinions expressed are solely those of the author(s) and do not necessarily reflect those of the European Union. Neither the European Union nor the European Commission can be held responsible for them.

\bibliographystyle{splncs04}
\bibliography{bib/bib}

@article{abraham2015,
  title   = {{GROMACS}: High performance molecular simulations through multi-level parallelism from laptops to supercomputers},
  journal = {SoftwareX},
  volume  = {1-2},
  pages   = {19-25},
  year    = {2015},
  issn    = {2352-7110},
  doi     = {10.1016/j.softx.2015.06.001},
  author  = {Mark James Abraham and Teemu Murtola and Roland Schulz and Szilárd Páll and Jeremy C. Smith and Berk Hess and Erik Lindahl}
}

@article{doi:10.1177/10943420231176527,
  author  = {Iker Martín-Álvarez and José I Aliaga and Maribel Castillo and Sergio Iserte and Rafael Mayo},
  title   = {Dynamic spawning of {MPI} processes applied to malleability},
  journal = {The International Journal of High Performance Computing Applications},
  volume  = {38},
  number  = {2},
  pages   = {69-93},
  year    = {2024},
  doi     = {10.1177/10943420231176527},
}

@misc{eupilot2021eu,
  title        = {EUPILOT: Pilot using Independent, Local and Open Technologies},
  howpublished = {\url{https://eupilot.eu/}},
  year         = {2021},
  note         = {Started December 2021 – Coordinated by Barcelona Supercomputing Center (BSC)}
}

@inproceedings{flexmpi,
  author    = {Mart\'{\i}n, Gonzalo and Marinescu, Maria-Cristina and Singh, David E. and Carretero, Jes\'{u}s},
  title     = {{FLEX-MPI}: an {MPI} extension for supporting dynamic load balancing on heterogeneous non-dedicated systems},
  year      = {2013},
  isbn      = {9783642400469},
  url       = {https://doi.org/10.1007/978-3-642-40047-6_16},
  doi       = {10.1007/978-3-642-40047-6_16},
  booktitle = {Proceedings of the 19th International Conference on Parallel Processing},
  pages     = {138–149},
  numpages  = {12},
  location  = {Aachen, Germany},
  series    = {Euro-Par'13}
}

@inproceedings{gabriel_2004,
  author    = {Gabriel, Edgar
               and Fagg, Graham E.
               and Bosilca, George
               and Angskun, Thara
               and Dongarra, Jack J.
               and Squyres, Jeffrey M.
               and Sahay, Vishal
               and Kambadur, Prabhanjan
               and Barrett, Brian
               and Lumsdaine, Andrew
               and Castain, Ralph H.
               and Daniel, David J.
               and Graham, Richard L.
               and Woodall, Timothy S.},
  title     = {Open {MPI}: Goals, Concept, and Design of a Next Generation {MPI} Implementation},
  booktitle = {Recent Advances in Parallel Virtual Machine and Message Passing Interface},
  year      = {2004},
  pages     = {97--104},
  isbn      = {978-3-540-30218-6},
  doi       = {10.1007/978-3-540-30218-6_19}
}

@article{Hess2008,
  author  = {Hess, Berk and Kutzner, Carsten and van der Spoel, David and Lindahl, Erik},
  title   = {GROMACS 4: Algorithms for Highly Efficient, Load-Balanced, and Scalable Molecular Simulation},
  journal = {Journal of Chemical Theory and Computation},
  volume  = {4},
  number  = {3},
  pages   = {435-447},
  year    = {2008},
  doi     = {10.1021/ct700301q}
}

@phdthesis{iserte_agut_high-throughput_2018,
  type    = {Ph.{D}. {Thesis}},
  title   = {High-throughput Computation through Efficient Resource Management},
  doi     = {10.6035/14101.2018.176272},
  urldate = {2025-08-07},
  school  = {Universitat Jaume I},
  author  = {Iserte, Sergio},
  month   = nov,
  year    = {2018}
}

@article{iserte_dmrlib_2020,
  title   = {{DMRlib}: Easy-coding and Efficient Resource Management for Job Malleability},
  volume  = {70},
  issn    = {0018-9340},
  doi     = {10.1109/TC.2020.3022933},
  journal = {IEEE Transactions on Computers},
  author  = {Iserte, S. and Mayo, R. and Quintana-Ortí, E. S. and Peña, A. J.},
  month   = sep,
  year    = {2020},
  pages   = {1443--1457}
}

@article{iserte_dynamic_2018,
  title   = {Dynamic Reconfiguration of Non-iterative Scientific Applications: A Case Study with {HPG}-aligner},
  volume  = {33},
  issn    = {1094-3420},
  doi     = {10.1177/1094342018802347},
  journal = {International Journal of High Performance Computing Application},
  author  = {Iserte, S. and Martínez, H. and Barrachina, S. and Castillo, M. and Mayo, R. and Peña, A. J.},
  month   = aug,
  year    = {2018},
  pages   = {1--10}
}

@inproceedings{iserte_malleable_2025,
  address   = {Merida, Yucatan, Mexico},
  title     = {Malleable Computational Fluid Dynamics Simulations},
  booktitle = {Proceedings of the 36th {Parallel} {CFD} {International} {Conference}},
  author    = {Iserte, S. and Houzeaux, G. and Sandås, P. and Peña, A. J. and Garcia-Gasulla, M.},
  month     = nov,
  year      = {2025}
}

@article{iserte_mpi_2025,
  title   = {{MPI} Malleability Validation under Replayed Real-World {HPC} Conditions},
  issn    = {0167-739X},
  doi     = {10.1016/j.future.2025.108305},
  urldate = {2025-12-14},
  journal = {Future Generation Computer Systems},
  author  = {Iserte, Sergio and Madon, Maël and Da Costa, Georges and Pierson, Jean-Marc and Peña, Antonio J.},
  month   = dec,
  year    = {2025},
  pages   = {108305}
}

@inproceedings{iserte_parallel_2024,
  address   = {Madrid, Spain},
  title     = {Parallel Efficiency-aware Standard MPI-based Malleability},
  booktitle = {Euro-par {Workshops} {Proceedings}},
  author    = {Iserte, Sergio and Lopez, Victor and Garcia-Gasulla, Marta and Peña, Antonio J.},
  month     = aug,
  year      = {2024}
}

@article{iserte_resource_2025,
  title   = {Resource optimization with {MPI} process malleability for dynamic workloads in {HPC} clusters},
  issn    = {0167-739X},
  doi     = {10.1016/j.future.2025.107949},
  journal = {Future Generation Computer Systems},
  author  = {Iserte, Sergio and Martín-Álvarez, Iker and Rojek, Krzysztof and Aliaga, José I. and Castillo, Maribel and Folwarska, Weronika and Peña, Antonio J.},
  year    = {2025},
  pages   = {107949}
}

@article{iserte_study_2020,
  title   = {A Study of the Effect of Process Malleability in the Energy Efficiency on GPU-based Clusters},
  volume  = {76},
  issn    = {0920-8542},
  doi     = {10.1007/s11227-019-03034-x},
  journal = {Journal of Supercomputing},
  author  = {Iserte, S. and Rojek, K.},
  month   = oct,
  year    = {2020},
  pages   = {255--274}
}

@inproceedings{leanmd,
  author    = {Prabhakaran, Suraj and Neumann, Marcel and Rinke, Sebastian and Wolf, Felix and Gupta, Abhishek and Kale, Laxmikant V.},
  booktitle = {IEEE International Parallel and Distributed Processing Symposium},
  title     = {A Batch System with Efficient Adaptive Scheduling for Malleable and Evolving Applications},
  year      = {2015},
  volume    = {},
  number    = {},
  pages     = {429-438},
  doi       = {10.1109/IPDPS.2015.34}
}

@inproceedings{lopez_talp_2021,
  author    = {Lopez, Victor and Ramirez Miranda, Guillem and Garcia-Gasulla, Marta},
  title     = {TALP: A Lightweight Tool to Unveil Parallel Efficiency of Large-scale Executions},
  year      = {2021},
  isbn      = {9781450383875},
  doi       = {10.1145/3452412.3462753},
  booktitle = {Proceedings of the 2021 on Performance EngineeRing, Modelling, Analysis, and VisualizatiOn STrategy},
  pages     = {3–10},
  numpages  = {8},
  location  = {Virtual Event, Sweden},
  series    = {PERMAVOST '21}
}

@article{Pronk2013,
  author  = {Pronk, Sander and Páll, Szilárd and Schulz, Roland and Larsson, Per and Bjelkmar, Pär and Apostolov, Rossen and Shirts, Michael R. and Smith, Jeremy C. and Kasson, Peter M. and van der Spoel, David and Hess, Berk and Lindahl, Erik},
  title   = {GROMACS 4.5: A high-throughput and highly parallel open source molecular simulation toolkit},
  journal = {Bioinformatics},
  volume  = {29},
  number  = {7},
  pages   = {845-854},
  year    = {2013},
  month   = {02},
  issn    = {1367-4803},
  doi     = {10.1093/bioinformatics/btt055}
}

@inproceedings{sudarsan-lammps,
  author    = {Sudarsan, Rajesh
               and Ribbens, Calvin J.
               and Farkas, Diana},
  title     = {Dynamic Resizing of Parallel Scientific Simulations: A Case Study Using {LAMMPS}},
  booktitle = {Computational Science (ICCS)},
  year      = {2009},
  pages     = {175--184},
  isbn      = {978-3-642-01970-8}
}

@inproceedings{sudarsan-nas,
  author    = {Sudarsan, Rajesh and Ribbens, Calvin J.},
  booktitle = {IEEE International Symposium on Parallel and Distributed Processing},
  title     = {Scheduling resizable parallel applications},
  year      = {2009},
  volume    = {},
  number    = {},
  pages     = {1-10},
  doi       = {10.1109/IPDPS.2009.5161077}
}

@misc{szilard_pall_2020_3893789,
  author    = {Szilárd Páll},
  title     = {Supplementary Information for Heterogeneous
               Parallelization and Acceleration of Molecular
               Dynamics Simulations in {GROMACS}
               },
  month     = jun,
  year      = 2020,
  publisher = {Zenodo},
  doi       = {10.5281/zenodo.3893789}
}

@article{tarraf_malleability_2024,
  title      = {Malleability in Modern {HPC} Systems: Current Experiences, Challenges, and Future Opportunities},
  issn       = {1558-2183},
  shorttitle = {Malleability in {Modern} {HPC} {Systems}},
  doi        = {10.1109/TPDS.2024.3406764},
  urldate    = {2024-05-31},
  journal    = {IEEE Transactions on Parallel and Distributed Systems},
  author     = {Tarraf, Ahmad and Schreiber, Martin and Cascajo, Alberto and Besnard, Jean-Baptiste and Vef, Marc-André and Huber, Dominik and Happ, Sonja and Brinkmann, André and Singh, David E. and Hoppe, Hans-Christian and Miranda, Alberto and Peña, Antonio J. and Machado, Rui and Gasulla, Marta Garcia- and Schulz, Martin and Carpenter, Paul and Pickartz, Simon and Rotaru, Tiberiu and Iserte, Sergio and Lopez, Victor and Ejarque, Jorge and Sirwani, Heena and Wolf, Felix},
  month      = jun,
  year       = {2024},
  pages      = {1--14},
}

@article{VanDerSpoel2005,
  author  = {Van Der Spoel, David and Lindahl, Erik and Hess, Berk and Groenhof, Gerrit and Mark, Alan E. and Berendsen, Herman J. C.},
  title   = {{GROMACS}: Fast, flexible, and free},
  journal = {Journal of Computational Chemistry},
  volume  = {26},
  number  = {16},
  pages   = {1701-1718},
  year    = {2005}
}

@inproceedings{yoo_slurm_2003,
  author    = {Yoo, Andy B.
               and Jette, Morris A.
               and Grondona, Mark},
  title     = {SLURM: Simple Linux Utility for Resource Management},
  booktitle = {Job Scheduling Strategies for Parallel Processing},
  year      = {2003},
  pages     = {44--60},
  isbn      = {978-3-540-39727-4},
  doi       = {10.1007/10968987_3}
}
\end{document}